\documentstyle[psfig,apjbib]{cupconf}

\ifoldfss
\else
  \ifnfssone
    \newmathalphabet{\mathit}
      \addtoversion{normal}{\mathit}{cmr}{m}{it}
      \addtoversion{bold}{\mathit}{cmr}{bx}{it}
    \newmathalphabet{\mathcal}
      \addtoversion{normal}{\mathcal}{cmsy}{m}{n}
    \else
    \ifnfsstwo
    \fi
  \fi
\fi

\def\CO#1{\ifnum#1=0                    % CO isotopes
           \ifmmode \mbox{\rm CO}
           \else {\rm CO}
           \fi
          \else
           \ifnum#1<15
            \ifmmode ^{#1}\mbox{\rm CO}
            \else $^{#1}${\rm CO}
            \fi
           \else
            \ifmmode \mbox{\rm C}^{#1}\mbox{\rm O}
            \else {\rm C}$^{#1}${\rm O}
            \fi
           \fi
          \fi}

\def\COp{\ifmmode \mbox{\rm CO}^+           %   +
         \else {\rm CO}$^+$                 % CO
         \fi}

\def\CS#1{\ifnum#1=0                    % CS isotopes
           \ifmmode \mbox{\rm CS}
           \else {\rm CS}
           \fi
          \else
           \ifnum#1<15
            \ifmmode ^{#1}\mbox{\rm CS}
            \else $^{#1}${\rm CS}
            \fi
           \else
            \ifmmode \mbox{\rm C}^{#1}\mbox{\rm S}
            \else {\rm C}$^{#1}${\rm S}
            \fi
           \fi
          \fi}

\def\HCOp{\ifmmode \mbox{\rm HCO}^+          %    +
          \else {\rm HCO}$^+$                % HCO
          \fi}
\def\Hthreep{\ifmmode \mbox{\rm H}_3^+         % H
             \else {\rm H}$_3^+$               %  3
             \fi}

\def\Htwo{\ifmmode \mbox{\rm H}_2              % H
          \else {\rm H}$_2$                    %  2
          \fi}

\def\HtwoO{\ifmmode \mbox{\rm H}_2\mbox{\rm O} % H O
           \else {\rm H}$_2${\rm O}            %  2
           \fi}

\def\ion#1#2{\ifmmode \mbox{{\rm #1}}\,\mbox{{\sc #2}} % generic ion
        \else {\rm #1}$\,${\sc #2}
        \fi}
\def\HI{\ion{H}{i}}

\def\rec#1#2{\if#2a                            % generic
              \ifmmode \mbox{{\rm #1}}\alpha   % recombination line
              \else {\rm #1}$\alpha$
              \fi
             \fi
             \if#2b
              \ifmmode \mbox{{\rm #1}}\beta
              \else {\rm #1}$\beta$
              \fi
             \fi
             \if#2g
              \ifmmode \mbox{{\rm #1}}\gamma
              \else {\rm #1}$\gamma$
              \fi
             \fi}

\def\Ha{\rec{H}{a}}                            % H alpha
\def\deg{\ifmmode ^\circ                % degrees symbol
         \else $^\circ$
         \fi
         \hskip -0.1truecm}
\def\degd#1.#2{                         % degrees over decimal point
               \ifmmode {#1^{\hskip 0.05em\circ}\hskip-0.42em.\hskip0.08em#2}
               \else {#1$^{\hskip 0.05em\circ}\hskip-0.42em.\hskip0.08em$#2}
               \fi
              }
\def\mind#1.#2{                         % minutes over decimal point
               \ifmmode {#1^{\hskip 0.05em\prime}\hskip-0.35em.\hskip0.05em#2}
               \else {#1$^{\hskip 0.05em\prime}\hskip-0.35em.\hskip0.05em$#2}
               \fi
              }
\def\secd#1.#2{                         % seconds over decimal point
               \ifmmode {#1^{\prime\prime}\hskip-0.46em.\hskip0.12em#2}
               \else {#1$^{\prime\prime}\hskip-0.46em.\hskip0.12em$#2}
               \fi
              }
\def\timsecd#1.#2{                      % time seconds over decimal point
                  \ifmmode {#1^{\rm s}\hskip-0.39em.\hskip0.08em#2}
                  \else {$#1^{\rm s}\hskip-0.39em.\hskip0.08em#2$}
                  \fi
                 }
\def\hms#1h#2m#3s{                      % hms format (for RA)
                  \relax
                  \ifmmode #1^{\rm h}\,#2^{\rm m}\,#3^{\rm s}
                  \else \hbox{$#1^{\rm h}\,#2^{\rm m}\,#3^{\rm s}$}
                  \fi
                 }
\def\dms#1d#2m#3s{                      % dms format (for Dec)
                  \relax
                  \ifmmode #1^\circ\,#2^{\prime}\,#3^{\prime\prime}
                  \else \hbox{$#1^\circ\,#2^{\prime}\,#3^{\prime\prime}$}
                  \fi
                 }
\def\dmsd#1d#2m#3.#4s{                  % dms format with decimal point (Dec)
                      \relax
                      \ifmmode #1^\circ\,#2^{\prime}\,#3^{\prime\prime}
                               \hskip-0.46em.\hskip0.12em#4
                      \else \hbox{$#1^\circ\,#2^{\prime}\,#3^{\prime\prime}
                            \hskip-0.46em.\hskip0.12em#4$}
                      \fi
                     }
\def\hm#1h#2m{                          % hm format (for RA)
              \relax
              \ifmmode #1^{rm h}\,#2^{\rm m}
              \else \hbox{$#1^{\rm h}\,#2^{\rm m}$}
              \fi
             }
\def\dm#1d#2m{                          % dm format (for Dec)
              \relax
              \ifmmode #1^\circ\,#2^{\prime}
              \else \hbox{$#1^\circ\,#2^{\prime}$}
              \fi
             }
\def\hmsd#1h#2m#3.#4s{                  % hms format with decimal point (RA)
                      \relax
                      \ifmmode #1^{\rm h}\,#2^{\rm m}\,#3^{\rm s}
                               \hskip-0.39em.\hskip0.08em#4
                      \else \hbox{$#1^{\rm h}\,#2^{\rm m}\,#3^{\rm s}
                            \hskip-0.39em.\hskip0.08em#4$}
                      \fi
                     }
\def\hmd#1h#2.#3m{                  % hm format with decimal point (RA)
                  \relax
                  \ifmmode #1^{\rm h}\,#2^{\rm m}
                           \hskip-0.55em.\hskip0.22em#3
                  \else \hbox{$#1^{\rm h}\,#2^{\rm m}
                        \hskip-0.55em.\hskip0.22em#3$}
                  \fi
                 }
\def\mg{\relax                          % magnitudes symbol
        \ifmmode ^{\rm m}
        \else $^{\rm m}$
        \fi
       }
\def\mgd#1.#2{                          % magnitudes over decimal point
              \relax
              \ifmmode #1^{\rm m}
                       \hskip-0.55em.\hskip0.22em#2
              \else \hbox{#1$^{\rm m}
                    \hskip-0.55em.\hskip0.22em$#2}
              \fi
             }

\def\la{\mathrel{\hbox{\rlap{\hbox{\lower4pt\hbox{$\sim$}}}\hbox{$<$}}}}
\def\ga{\mathrel{\hbox{\rlap{\hbox{\lower4pt\hbox{$\sim$}}}\hbox{$>$}}}}

\def\unitspace{\;}                      % space to be used with units

\def\un#1{\ifmmode \unitspace\mbox{\rm #1} % generic unit
          \else $\unitspace$#1
          \fi}
\def\pun#1#2{\ifmmode \unitspace\mbox{\rm #1}^{#2} % generic unit with a power
             \else $\unitspace$#1$^{#2}$
             \fi}

\def\kms{\un{km}\pun{s}{-1}}          %     -1
\def\Lsun{\ifmmode \un{L}_{\odot}     % solar luminosity
          \else $\un{L}_{\odot}$
          \fi}
\def\Msun{\ifmmode \un{M}_{\odot}     % solar mass
          \else $\un{M}_{\odot}$
          \fi}
\def\mum{\ifmmode \unitspace\mu\mbox{\rm m} % micron
         \else $\unitspace\mu$m
         \fi}
\def\pyr{\pun{yr}{-1}}    %   -1
\def\Bp{\relax                            % B
        \ifmmode B_{||}                   %  ||
        \else $B_{||}$
        \fi}
\def\Bt{\relax                            % B
        \ifmmode B\!_{\perp}              %  |
        \else $B\!_{\perp}$               %  -
        \fi}
\def\Gcr{\relax                           % Gamma
         \ifmmode \Gamma\!_{\rm cr}       %      cr
         \else $\Gamma\!_{\rm cr}$
         \fi}
\def\ICII{\relax                          % I
          \ifmmode I_{[\CII]}             %  [C II]
          \else $I_{[\CII]}$
          \fi}
\def\LHtwo{\relax                                 % L
           \ifmmode L_{\mbox{\rm\scriptsize H}_2} %  H
           \else $L_{\mbox{\rm\scriptsize H}_2}$  %   2
           \fi}
\def\LLya{\relax                          % L
          \ifmmode L_{{\rm Ly}\,\alpha}   %  Ly alpha
          \else $L_{{\rm Ly}\,\alpha}$
          \fi}
\def\MHtwo{\relax                                 % M
           \ifmmode M_{\mbox{\rm\scriptsize H}_2} %  H
           \else $M_{\mbox{\rm\scriptsize H}_2}$  %   2
           \fi}
\def\MHtwodot{\relax                                       % .
              \ifmmode \dot{M}_{\mbox{\rm\scriptsize H}_2} % M
              \else $\dot{M}_{\mbox{\rm\scriptsize H}_2}$  %  H
              \fi}                                         %   2
\def\Mstardot{\relax                      % .
              \ifmmode \dot{M}_{\ast}     % M
              \else $\dot{M}_{\ast}$      %  *
              \fi}
\def\nHI{\relax                                      % n
         \ifmmode n_{\mbox{\scriptsize\rm H\,\sc I}} %  HI
         \else $n_{\mbox{\scriptsize\rm H\,\sc I}}$
         \fi}
\def\nHtwo{\relax                                % n
           \ifmmode n_{{\mbox{\scriptsize H}}_2} %  H
           \else $n_{{\mbox{\scriptsize H}}_2}$  %   2
           \fi}
\def\rhostardot{\relax                         %  .
                \ifmmode \dot{\rho}_{\ast}     % rho
                \else $\dot{\rho}_{\ast}$      %    *
                \fi}
\def\rhoZdot{\relax                          %  .
             \ifmmode \dot{\rho}_{\rm Z}     % rho
             \else $\dot{\rho}_{\rm Z}$      %    Z
             \fi}

\def\sou#1#2{\relax                       % source designations
             \ifmmode {\rm #1}\,{\rm #2}  % e.g. \sou{Arp}{220}
             \else #1$\,$#2
             \fi}

\def\NGC#1{\sou{NGC}{#1}}                % NGC number

\def\Arp#1{\sou{Arp}{#1}}                % Arp number

\def\qu#1#2{\relax                          % quantity symbols
            \ifmmode #1_{\rm #2}            % e.g. \qu{L}{FIR}
            \else $#1_{\rm #2}$
            \fi}

\newcommand{\figref}[1]{Fig.~\protect\ref{#1}}
\newcommand{\twoeqsref}[2]{Eqs.~$\left(\protect\ref{#1}\right)$ and $\left(\protect\ref{#2}\right)$}
\newcommand{\secref}[1]{\S\,\ref{#1}}

\def\hexnumber#1{\ifcase#1 0\or1\or2\or3\or4\or5\or6\or7\or8\or9\or
 A\or B\or C\or D\or E\or F\fi }

\makeatletter
\ifx\CUP@mtlplain@loaded\undefined
\else
\fi
\makeatother
\makeatletter
\ifx\CUP@mtlplain@loaded\undefined
\else
\fi
\makeatother
\title[H$_2$ emission in starforming galaxies]{%
H$_2$ emission as a diagnostic of physical processes in starforming galaxies}

\author[P.P.\ van der Werf]{%
Paul\ns
P.\ns
V\ls A\ls N\ns
D\ls E\ls R\ns
W\ls E\ls R\ls F$^1$}

\affiliation{$^1$Leiden Observatory, P.O.\ Box 9513, NL~-~2300~RA
Leiden, The Netherlands}

\begin{document}

\maketitle

\makeatletter
\renewcommand{\@makefnmark}{\mbox{\ }}
\makeatother

\renewcommand{\thefootnote}{}

\footnote{\noindent to appear in {\it
``Molecular hydrogen in space''}, eds.\ F.\ Combes
\& G.\ Pineau des For\^ets, Cambridge University Press}

\ifnfssone
\else
  \ifnfsstwo
  \else
    \ifoldfss
      \let\mathcal\cal
      \let\mathrm\rm
      \let\mathsf\sf
    \fi
  \fi
\fi

\begin{abstract}
Observations and interpretation of extragalactic rotational and
rovibrational  $\Htwo$ emission are reviewed. Direct observations of
$\Htwo$ lines do not trace bulk $\Htwo$ mass, but excitation
rate. As such, the $\Htwo$ lines are unique diagnostics, if the
excitation mechanism can be determined, which generally requires high-quality
spectroscopy and suitable additional data. The diagnostic power of the 
$\Htwo$ lines is illustrated by two cases studies: $\Htwo$ purely
rotational line emission from the disk of the nearby spiral galaxy
$\NGC{891}$ and high resolution imaging and spectroscopy of $\Htwo$
vibrational line emision from the luminous merger $\NGC{6240}$. 
\end{abstract}

\firstsection % if your document starts with a section,
\section{Introduction}
\label{sec.introduction}

Direct observations of $\Htwo$ emission from external galaxies have
become standard practice in the past decade through the revolution in
ground-based near-infrared instrumentation. 
As a result, the
near-infrared $\Htwo$ rovibrational lines are now readily detectable
throughout the local universe
(e.g., Moorwood \& Oliva
1988, 1990; Puxley {\it et al.} 1988, 1990; Goldader {\it et al.}
1995, 1997; \citebare{Vanzietal98}).
More recently, the Short Wavelength
Spectrograph (SWS) 
on the Infrared Space Observatory (ISO) has for the 
first time allowed detection of the purely rotational $\Htwo$ lines in 
the mid-infrared spectral regime. For instance, the first detection
(outside the solar system) of the $\Htwo$ S(0) line at $28.21\mum$ was
reported by \citetext{Valentijnetal96} from the star forming
nucleus of the nearby spiral galaxy $\NGC{6946}$.
\nocite{MoorwoodOliva88}
\nocite{MoorwoodOliva90}
\nocite{Puxleyetal88}
\nocite{Puxleyetal90}
\nocite{Goldaderetal95}
\nocite{Goldaderetal97}

The interpretation of these data is, however, far from trivial. At
typical molecular cloud temperatures ($T\sim20\un{K}$) 
the upper levels of even the
lowest $\Htwo$ transitions are essentially unpopulated, and hence
$\Htwo$ emission is, unlike for instance CO $J=1{\to}0$ emission, not
a tracer of bulk molecular gas mass. Instead, an excitation
mechanism capable of populating these energy levels is required. 
Furthermore, the
excitation rate needs to be high enough to maintain an excited 
level population sufficient for producing detectable emission.
Hence the $\Htwo$ line luminosities measure the {\it rate
of excitation\/}; as such they provide unique and highly diagnostic
information, that cannot be obtained in any other way. In addition,
since the $\Htwo$ lines are forbidden quadrupole transitions with very
small Einstein $A$ values, the lines are optically thin; 
hence there is, in contrast to the situation with most other molecular
lines, no need to solve a
complicated, geometry-dependent radiative transfer problem for a
physical interpretation of the $\Htwo$ lines. The near-infrared
$\Htwo$ lines will of course suffer from extinction by dust, but this
effect can usually be quantified adequately by using suitable ratios
of hydrogen recombination lines, [$\ion{Fe}{ii}$] lines or $\Htwo$ 
rovibrational lines in the same spectral range, with accurately known
intrinsic flux ratios.

The situation is complicated, however, by the fact that $\Htwo$
excitation can be brought about by a variety of mechanisms, which may
all play a role. For instance, in a starburst galaxy, $\Htwo$
emission may be generated by UV-pumping in starforming regions or by
shocks due to supernova remnants or outflows; moreover, these
processes are expected to occur together in the same small volume
occupied by the starburst, and 
thus a combination of excitation mechanisms may be expected at every
position. Generally, this combination will be difficult to separate 
(e.g., $\NGC{253}$ observations:
\citebare{Forbesetal93}; \citebare{Engelbrachtetal98}). 
In addition, shocks due
to large-scale streaming motions in spiral arms, bars (favoured for
the barred Seyfert~2
$\NGC{1068}$ by \citebare{Tacconietal94}) or merger-driven 
flows (e.g., in $\NGC{6240}$, \citebare{VanDerWerfetal93}; see also
\secref{sec.NGC6240}) may play a role. 
Furthermore, X-ray excitation may be produced by multiple supernova
remnants, or, if present, by an active galactic nucleus (proposed for
e.g., Cyg~A by \citebare{Wardetal91} and for
Cen~A by \citebare{BryantHunstead99}) or a cooling flow
\cite{JaffeBremer97}. In the absence of sufficient spatial resolution
for separating the various excitation mechanisms, spectral diagnostics 
must be used. Models for UV-pumped, shocked or X-ray excited $\Htwo$
emission have reached considerable predictive power; however, the
densities of the emitting regions are often sufficiently high to
thermalize the relevant level populations, quenching the typical
signatures of non-thermal excitation processes. The only accessible
diagnostics are then fluxes of lines with very high critical
densities. However, these lines are intrinsically faint, and 
accurate fluxes for such lines require long integrations with large 
telescopes. 
The complexity of the problem is well illustrated by the $\Htwo$
emission of the merging system $\NGC{6240}$, which has been attributed 
to X-ray excitation \cite{DraineWoods90}, UV-pumping \cite{Tanakaetal91},
shocks \cite{VanDerWerfetal93} and
formation pumping \cite{MouriTaniguchi95}. However, all of the
non-thermal excitation processes relied on poorly measured fluxes of
faint lines. More recent spectroscopy of $\NGC{6240}$ by
\citetext{Sugaietal97} showed that only shock excitation can account
for the $\Htwo$ rovibrational spectrum (see also
\secref{sec.NGC6240}).

Generally a combination of accurate multi-line spectroscopy and
suitable additional information such as high resolution spatial
information is needed for a proper analysis of the
dominant excitation mechanism and hence a physical analysis of the
$\Htwo$ emission. A complicating factor when combining $\Htwo$
rovibrational and purely rotational lines is the fact that the dominant
excitation mechanisms of these lines may be different, because of the
different excitation requirements of these transitions: while the lowest
rotational lines require $T>100\un{K}$ and are excited at moderate
densities ($n_{{\mathrm H}_2}>10^2\pun{cm}{-3}$), the rovibrational
lines require $T>2000\un{K}$ and $n_{{\mathrm H}_2}>10^4\pun{cm}{-3}$.
The following sections will therefore present two case studies. First
an analysis of the extended purely rotational $\Htwo$ emission from
the nearby edge-on spiral galaxy $\NGC{891}$ is given
(\secref{sec.NGC891}). Then, the more extreme conditions in the merger
$\NGC{6240}$ are discussed, using new high resolution near-infrared
data. The general implications of these two cases are discussed
in \secref{sec.implications}.

\section{Extended H$_2$ rotational line emission in the disk of NGC\,891}
\label{sec.NGC891}

\citetext{ValentijnVanDerWerf99b} have observed
seven positions in the disk of
the nearby (distance $D=9.5\un{Mpc}$) edge-on
spiral galaxy $\NGC{891}$ with the ISO SWS\null.
The positions observed include the nucleus and
positions spaced along the disk to galactocentric distances $R$ of
$8\un{kpc}$ south of the nucleus and $11\un{kpc}$ north of the
nucleus. At the $11\un{kpc}$ north position, the $\CO{12}$ $J=1{\to}0$ 
\cite{GarciaBurilloetal92,Scovilleetal93,Sakamotoetal97}, and 
dust emission \cite{Israeletal99} are barely detected.
With ISO, the $\Htwo$ S(0) ($28.21\mum$) and S(1) ($17.03\mum$) lines were
detected at all positions. These are the first detections of these
lines outside starburst or active nuclei, with the exception of the
detection in the disk of $\NGC{6946}$ reported by
\citetext{ValentijnVanDerWerf99a}.

\begin{figure} 
\centerline{\psfig{figure=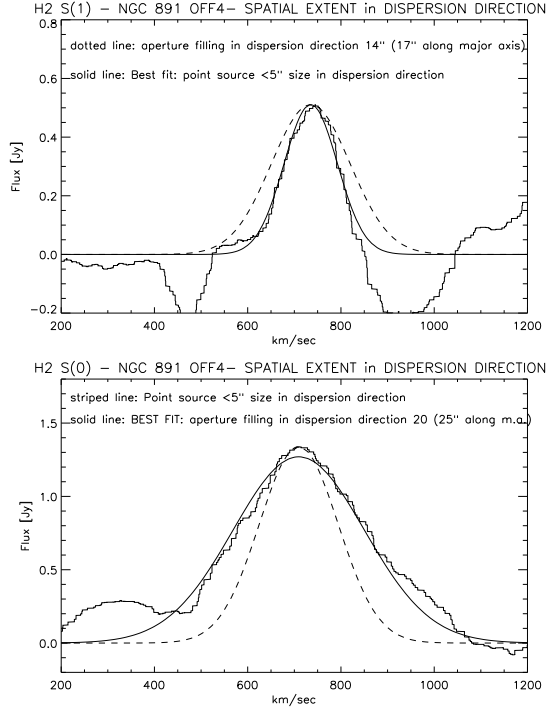,width=7.4cm}}    
\caption{$\Htwo$ S(1) (upper panel) and S(0) (lower panel) lines from
$R=8\un{kpc}$ south in the disk of $\NGC{891}$, with fits for a point
source and an aperture-filling source as indicated.}
\label{fig.NGC8918kpcS}
\end{figure} 

The simplest analysis of these data would assume 
that the emission is fully thermalized
(i.e., the high-density limit is assumed) and arises
from an isothermal gas layer with an ortho-para
ratio of three, in agreement with the statistical weights of the ortho-
and para-varieties.
Under these assumptions the S(1)/S(0) ratio yields the
temperature of the emitting region, which can be combined with the
observed fluxes to give the column density of emitting $\Htwo$,
averaged over the SWS aperture. However, these simple assumptions lead
to unacceptable results. For instance, at the position
$8\un{kpc}$ north (a typical ``disk'' position), 
the S(0) and S(1) data then imply $N(\Htwo)=2.7\cdot10^{22}$ 
at $T=76\un{K}$. This result is incompatible with the CO
$J=2{\to}1$/$J=1{\to}0$ line ratio of 0.75, at this position 
\cite{GarciaBurilloetal92}, which strongly rules out a large mass of
dense molecular gas at $T=76\un{K}$. This argument shows that the
simple assumptions used above do not suffice.

The discrepancy can be removed by allowing lower densities, and
therefore subthermal excitation of the $\Htwo$ $J{=}3$ level, giving 
rise to a higher implied temperature for a given S(1)/S(0)
ratio. Because the temperatures involved are still much lower than the 
upper levels of the transitions involved, the implied $\Htwo$ mass
will be extremely sensitive to the adopted temperature, and thus
strongly decrease as lower densities are allowed. An additional
advantage of this procedure is that the CO $J{=}2$ level will now also be
subthermally populated, so that CO $J=2{\to}1$/$J=1{\to}0$ line ratios 
lower than the thermal value can be tolerated in the analysis. Indeed
the multi-level multi-isotope CO data by
\citetext{GarciaBurilloetal92} indicate dominant densities of
$n_{{\mathrm H}_2}\sim10^3\pun{cm}{-3}$ in the nuclear region and
$n_{{\mathrm H}_2}\sim200\pun{cm}{-3}$ throughout the disk of
$\NGC{891}$. Adopting these densities, the ISO data imply 
$\Htwo$ column densities
varying from 10 to 30\% of the value derived from CO, at temperatures of 
120 to $130\un{K}$, throughout the inner disk ($R<8\un{kpc}$) of $\NGC{891}$. 

At the outer disk positions ($R=8-11\un{kpc}$) however, this solution
is not adequate, since the S(0) and S(1) line profiles
at these positions are
significantly different. The SWS uses an aperture much larger than the 
diffraction beam and hence the observed line width depends on the
extent of the emission region; the S(0) line shows the broad profile
characteristic of aperture-filling emission, while the S(1) profile
is narrow, indicating emission from a region much smaller than the
aperture (\figref{fig.NGC8918kpcS}). Thus the S(0) and S(1) lines in
the outher disk of $\NGC{891}$ must originate in separate regions: a
warm component ($T>130\un{K}$) 
dominating the S(1) emission and located in isolated
regions in the disk and a separate more pervasive component, 
dominating the S(0) line. The latter component may contain very significant
mass, depending on its temperature, 
and the implications of this possibility for the mass budget at
the outer positions have been explored by \citetext{ValentijnVanDerWerf99b}.
However, a solution where the component dominating the S(0) line is
cool ($T\sim90\un{K}$) is problematic, since the heating required for
maintaining a very large mass of $\Htwo$ at $T\sim90\un{K}$ cannot
be provided. In thermal equilibrium the heating rate should 
equal the cooling rate, which is dominated by [$\ion{C}{ii}$]
$158\mum$ emission (which at $T\sim90\un{K}$ is an extremely efficient 
cooler, since the upper level of the $158\mum$ transition is at
$91\un{K}$), with possible contributions from $\Htwo$
rotational lines, CO rotational lines, and [$\ion{C}{i}$]
emission. The [$\ion{C}{ii}$] $158\mum$ emission from $\NGC{891}$ has
been observed by \citetext{Maddenetal94}, and it is easily verified
that in the outer disk the [$\ion{C}{ii}$] emission dominates the
$\Htwo$ emission (SWS data discussed here), CO emission (barely
detected in the outer disk) and [$\ion{C}{i}$] emission
\cite{GerinPhillips97}. If all available carbon (assumed to have solar
abundance) is in the form of C$^+$, the [$\ion{C}{ii}$] emission
allows a column density of at most
$N(\Htwo)\sim7\cdot10^{21}\pun{cm}{-2}$ at $T\sim90\un{K}$ 
in the outer disk, averaged over the SWS aperture, while 
$N(\Htwo)\sim10^{23}\pun{cm}{-2}$ would be required to produce the
observed S(0) line flux.

This problem can be solved by relaxing the final assumption: that of
an ortho-para of three. A lower ortho-para ratio raises the
implied temperature and lowers the implied mass in the same way as a
lower density does (as discussed above). Assuming an ortho-para ratio
of unity, the implied mass is lowered sufficiently that the required
heating can be accounted for.

This analysis thus favours an interpretation in which the $\Htwo$
rotational emission arises in low-density 
($n_{{\mathrm H}_2}\sim200\pun{cm}{-3}$), warm ($T\sim120\un{K}$)
molecular gas with an ortho-para ratio of about unity, which pervades
the disk of $\NGC{891}$ at least to the end of the detectable CO disk
and dominates the S(0) emission; it contains 10 to 30\% of the mass
implied by CO observations (lower fractions will result if an
ortho-para ratio below unity is adopted);
concentrations of warmer gas (plausibly identified with active star
forming regions) are ubiquitous in the inner disk and rarer in the
outer disk and give rise to the S(1) emission. Given the
density and temperature of the extended warm gas, this component can
most likely be identified with extended low-density photon-dominated regions
(PDRs) which form the warm envelopes of giant molecular clouds,
heated by the local interstellar radiation field. Observations of
Galactic molecular cloud edges in the $21\un{cm}$ $\HI$ line reveal the
presence of such warm envelopes by a ``limb brightening'' in $\HI$ 
(\citebare{Wannieretal83}; Van der Werf {\it et al.}\ 1988, 1989).
Modeling of molecular cloud
envelopes by \citetext{AnderssonWannier93} shows that the temperatures 
and densities estimated here are typical for such regions,
especially in the zone where the transition from molecular to atomic 
hydrogen takes place. This interpretation of the present $\Htwo$ data
in terms of extended diffuse
PDRs in cloud envelopes is corroborated by the excellent
numerical agreement with column density estimates based on
[$\ion{C}{ii}$] $158\mum$, the principal coolant for such regions. In
addition, the fact that the implied ortho-para ratio is significantly
lower than three also points to a low-density PDR origin
of this emission, since shocks or hot, high-density PDRs would produce 
an ortho-para ratio of three (the high temperature thermal value) 
by spin exchange reactions with H and H$^+$. In summary, the $\Htwo$
rotational lines in the disk of $\NGC{891}$ arise in  warm, extended 
molecular cloud envelopes; these envelopes provide a physical link
between the giant molecular clouds and the atomic medium in which they 
are embedded \cite{Chromeyetal89}. The $\Htwo$ rotational lines provide 
unique diagnostics of these regions.
\nocite{VanDerWerfetal88}
\nocite{VanDerWerfetal89}

\section{Vibrational H$_2$ emission in the nuclear region of the
  luminous merger NGC\,6240}
\label{sec.NGC6240}

The vibrational $\Htwo$ emission of $\NGC{6240}$ has attracted
attention because of its extraordinary luminosity: $7\cdot10^7\Lsun$
in the $\Htwo$ $v=1{\to}0$ S(1) line alone (for
$H_0=75\kms\pun{Mpc}{-1}$ and with no correction for extinction). 
This line contains 0.012\% of the
bolometric luminosity of $\NGC{6240}$, which is considerably higher 
than any other galaxy \cite{VanDerWerfetal93}. Together the
vibrational lines may account for 0.1\% of the total bolometric luminosity.

\begin{figure} 
\centerline{\psfig{figure=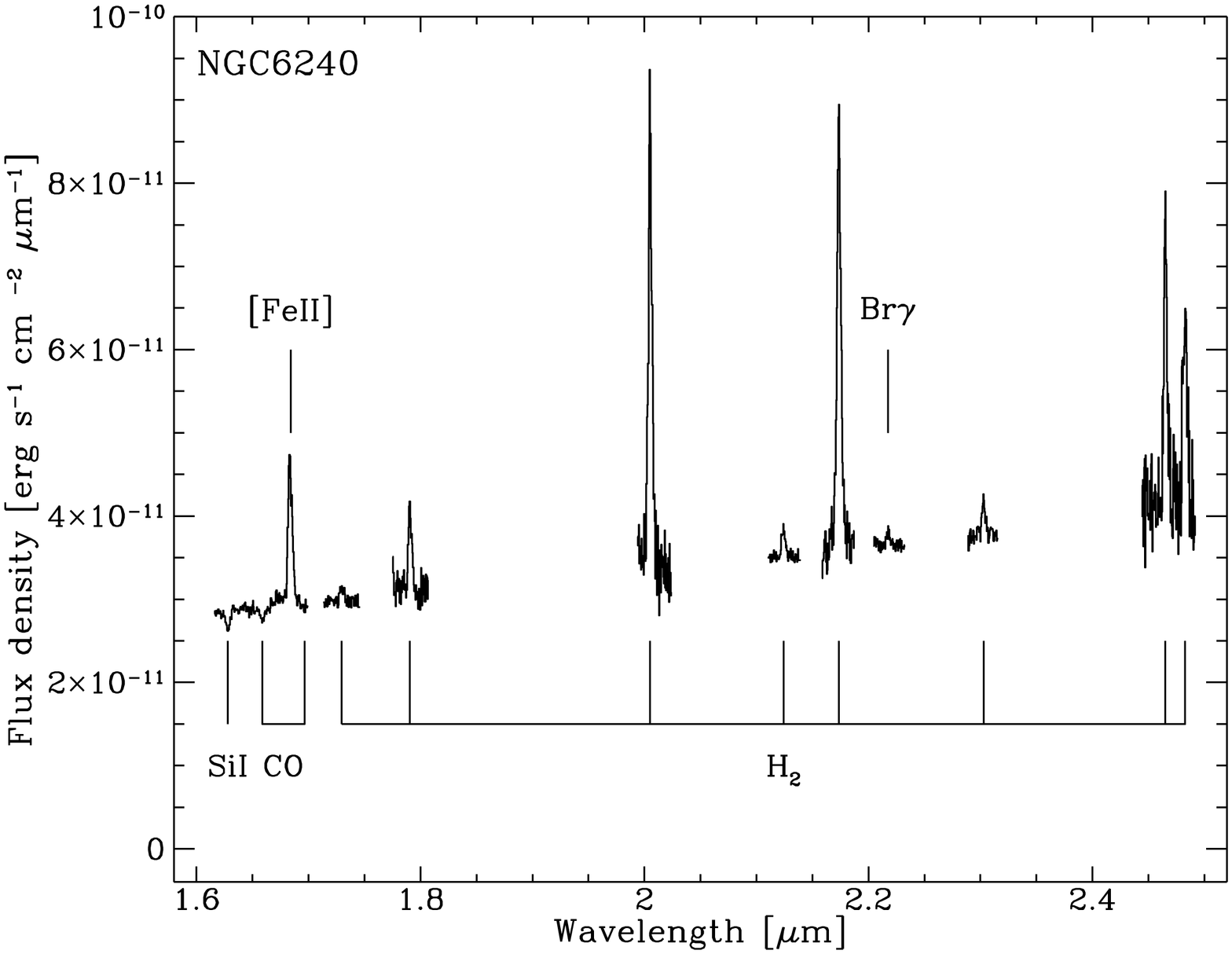,width=11.2cm,angle=-90}}    
\caption{$H{+}K$-band spectrum of $\NGC{6240}$ showing $\Htwo$ lines
  up to the $v=1{\to}0$ S(9) line \protect\cite{VanDerWerfetal00}.}
\label{fig.NGC6240H2spectrum}
\end{figure}

Imaging of the $\Htwo$ $v=1{\to}0$ S(1) emission from $\NGC{6240}$
has shown that the $\Htwo$ emission peaks {\it between\/} the two
remnant nuclei of the merging system \cite{VanDerWerfetal93}. 
This morphology provides a
unique constraint on the excititation mechanism, since it argues
against any scenario where the excitation is dominated by the stellar
component (e.g., UV-pumping, excitation by shocks or X-rays from
supernova remnants). Instead, the favoured excitation mechanism is
slow shocks in the nuclear gas component, which, as shown by 
recent high resolution interferometry in the CO 
$J=2{\to}1$ line \cite{Tacconietal99}, 
also peaks between the nuclei of $\NGC{6240}$. 
A multi-line $\Htwo$
vibrational spectrum (\figref{fig.NGC6240H2spectrum}) confirms this 
excitation mechanism \cite{VanDerWerfetal00}, in agreement with
more limited spectroscopy by \citetext{Sugaietal97}. The 
interpretation in terms of slow shocks also naturally accounts for the 
high ratio of $\Htwo$ line emission to infrared continuum emission,
which is a characteristic of such shocks (e.g., \citebare{Draineetal83}).

What is the role of these shocks? In the shocks mechanical
energy is dissipated and radiated away, mostly in spectral lines
(principally $\Htwo$, CO, $\HtwoO$ and [$\ion{O}{i}$] lines). This
energy is radiated away at the expense of the
orbital energy of the molecular clouds in the central potential well. 
Consequently, the dissipation of mechanical energy by the shocks will
give rise to an infall of molecular gas to the centre of the potential 
well. Therefore,
{\it the $\Htwo$ vibrational lines measure the rate of infall of
  molecular gas\/} into the central potential well in $\NGC{6240}$. 
This conclusion can be quantified by writing
\begin{equation}
\qu{L}{rad}=\qu{L}{dis},
\label{eq.L}
\end{equation}
where $\qu{L}{rad}$ is the total luminosity radiated by the shocks and
$\qu{L}{dis}$ the dissipation rate of mechanical energy, giving rise
to a molecular gas infall rate $\dot{M}_{{\rm H}_2}$ given by
\begin{equation}
\qu{L}{dis}={1\over2}\dot{M}_{{\rm H}_2}v^2,
\label{eq.MH2dot}
\end{equation}
where $v$ is the circular
orbital velocity at the position where the shock occurs.

Using a $K$-band extinction of $\mgd 0.15$ \cite{VanDerWerfetal93},
the total luminosity of $\Htwo$ vibrational lines from $\NGC{6240}$
becomes $7.2\cdot10^8\Lsun$; inclusion of the purely rotational lines
observed with the ISO SWS approximately doubles this number, so that
$\qu{L}{rad}=1.5\cdot10^9\Lsun$. 

\begin{figure} 
\centerline{\psfig{figure=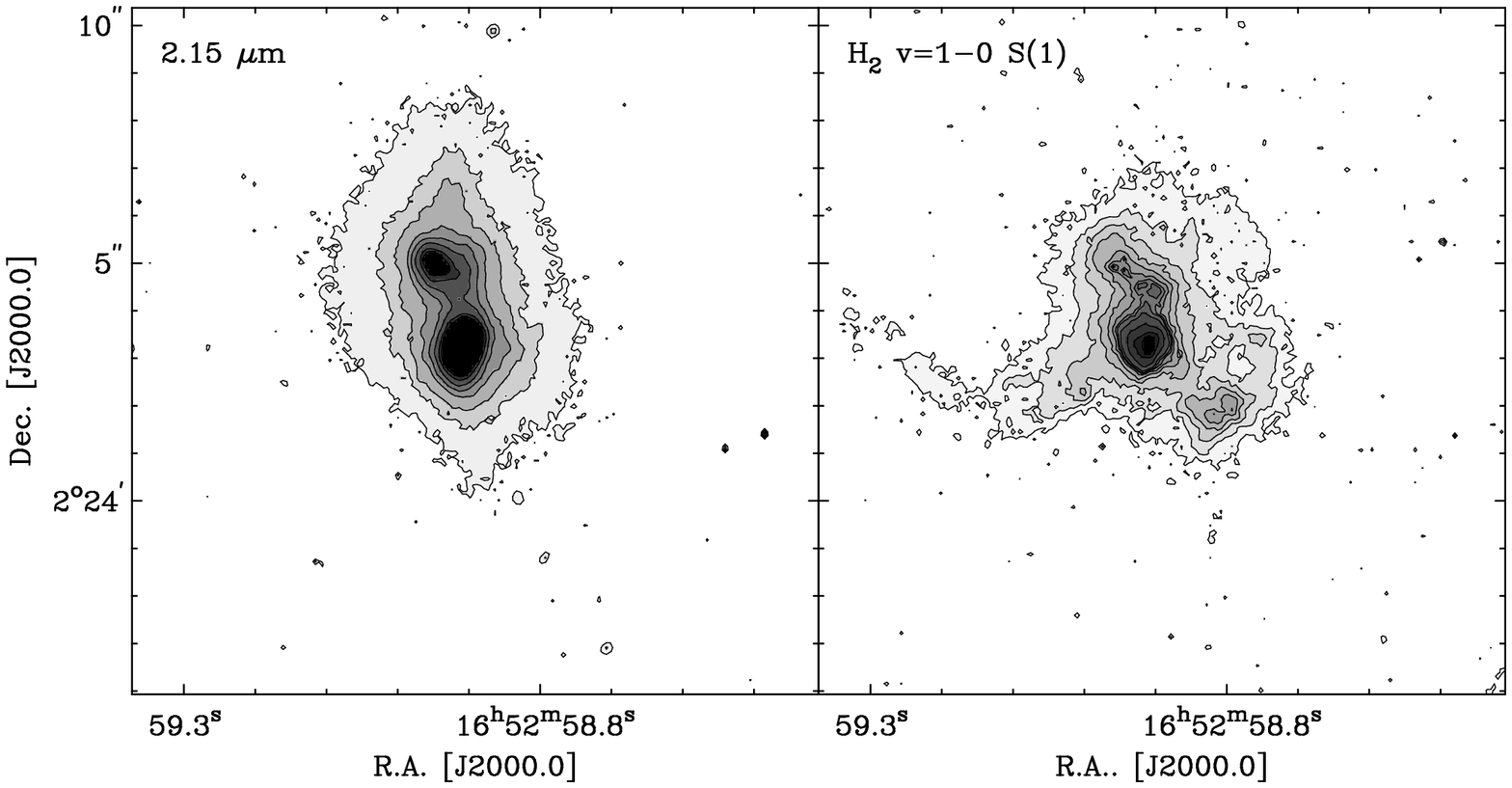,width=13.5cm,angle=-90}}    
\caption{High resolution imaging of $\NGC{6240}$ in the $\Htwo$
  $v=1{\to}0$ S(1) line and the $2.15\mum$ continuum with NICMOS/HST \protect\cite{VanDerWerfetal00}.}
\label{fig.NGC6240NICMOS}
\end{figure}

In order to use this number to estimate $\dot{M}_{{\rm H}_2}$, it is
necessary to establish more accurately the fraction of the $\Htwo$
emission that is due to infalling gas. Observations with NICMOS on the 
Hubble Space Telescope (HST) by \citetext{VanDerWerfetal00} 
provide the required information
(\figref{fig.NGC6240NICMOS}). The NICMOS image shows that the emission
consists of a number of tails (presumably related to the superwind
also observed in $\Ha$ emission), and concentrations assocated with
the two nuclei, and a further concentration approximately (but not
precisely) between the two nuclei. The relative brightness of the
$\Htwo$ emission from the southern nucleus is deceptive, since this
nucleus is much better centred in the filter that was used for these
observations than the other emission components, in particular the
northern nucleus. Taking this effect into account, it is found that
32\% of the total $\Htwo$ flux is associated with the southern
nucleus, 16\% is associated with the northern nucleus, and 12\% with
the component between the two nuclei, the remaining 40\% being
associated with extended emission. Using inclination-corrected
circular velocities (from \citebare{Tecza99}) of $270$ and $360\,\kms$ 
for the southern en northern nucleus respectively, and of $280\kms$
for the central component \cite{Tacconietal99}, the mass infall rates 
derived using \twoeqsref{eq.L}{eq.MH2dot} are $80\Msun\pun{yr}{-1}$
for the southern nucleus, $22\Msun\pun{yr}{-1}$ for the northern
nucleus and $28\Msun\pun{yr}{-1}$ for the central component.

The derived
molecular gas inflow rate to the two nuclei is remarkably close to the 
mass consumption rate by star formation of approximately $60\Msun\pyr$, 
indicating that the $\Htwo$ emission from the nuclei directly measures 
the fueling of the starbursts in these regions. The gas inflow towards 
the central component is more remarkable, since this component is not 
associated with a prominent stellar component. 
The gravitational potential at this position is therefore most likely
dominated by the gas itself, which is also indicated by high
resolution 
interferometry in the CO $J=2{\to}1$ line \cite{Tacconietal99}. The absence of
prominent star formation at this position shows that the central gas
component is gravitationally stabilized, possibly by a high local
shear. However, as the central gas column density increases, the shear
must eventually be overcome and given the high gas density (and hence
short free-fall time) an explosive starburst will result. In that
stage $\NGC{6240}$ will become a true ultraluminous infrared galaxy.

\section{General implications} 
\label{sec.implications}

The two case studies discussed above illustrate the unique diagnostic power
of $\Htwo$ emission lines provided that the excitation mechanism can
be determined. The extended PDR emission detected in $\NGC{891}$
reveals the presence of a widespread fluorescent component, which
is probably a common feature of star forming galaxies. For instance,
\citetext{Paketal96} have detected extended diffuse $\Htwo$
vibrational emission from the inner Milky Way and argue that this
emission is UV-excited; similarly, \citetext{Harrisonetal98} argue for 
purely fluorescent extended $\Htwo$ emission
in the moderate luminosity starburst galaxy $\NGC{253}$. 

The configuration of shocked $\Htwo$ emission from a pronounced
molecular gas concentration between the two nuclei as found in
$\NGC{6240}$ is not unique either: it is also 
found in $\Arp{220}$ \cite{VanDerWerfIsrael00},
and several other mergers. Perhaps the best case in point
is the well-known merging ``Antennae'' system ($\NGC{4038-4039}$)
where pronounced CO emission is found in the interaction zone between
the two nuclei \cite{Stanfordetal90}, which is the site of vigorous
obscured star formation \cite{Mirabeletal98}. These geometries may be 
due to the fact that the gas components are dissipative, and thus
merge on a shorter time scale than the dissipationless stellar
components, which merge by the slower process of dynamical friction. 
The $\Htwo$ emission thus provides unique insight into the physics of
these gas-rich mergers.

\begin{acknowledgments}
I would like to thank my collaborators Frank Israel, Alan Moorwood,
Tino Oliva,
and Edwin Valentijn for discussions on this subject, and Guido Kosters 
for reducing the NICMOS data of $\NGC{6240}$.
\end{acknowledgments}

\bibliographystyle{astrobib_apj}

\bibliography{%
strings,%
Arp220,%
CenA,%
coolingflows,%
CygA,%
nearIR,%
farIR,%
GalacticCentre,%
HIdarkclouds,%
L134,%
L1551,%
M82,%
NGC253,%
NGC1068,%
NGC4038-39,%
NGC6240,%
NGC6946,%
NGC891,%
OrionA,%
Seyferts,%
shocks,%
starbursts,%
ULIGs,%
Xrayexcitation%
}

\end{document}